# Three-dimensional droplets of swirling superfluids

Yaroslav V. Kartashov,[1,2] Boris A. Malomed,[3,4] Leticia Tarruell[1], and Lluis Torner[1,5]

[1]ICFO-Institut de Ciencies Fotoniques, The Barcelona Institute of Science and Technology, 08860 Castelldefels (Barcelona), Spain
[2]Institute of Spectroscopy, Russian Academy of Sciences, Troitsk, Moscow, 108840, Russia
[3]Department of Physical Electronics, School of Electrical Engineering, Faculty of Engineering, and Center for Light-Matter Interaction,Tel Aviv University, 69978 Tel Aviv, Israel
[4]ITMO University, St. Petersburg 197101, Russia
[5]Universitat Politecnica de Catalunya, 08034, Barcelona, Spain
*Corresponding author: Yaroslav.Kartashov@icfo.eu

## Abstract

A new method for the creation of 3D solitary topological modes, corresponding to vortical droplets of a two-component dilute superfluid, is presented. We use the recently introduced system of nonlinearly coupled Gross-Pitaevskii equations, which include contact attraction between the components, and quartic repulsion stemming from the Lee-Huang-Yang correction to the mean-field energy. Self-trapped vortex tori, carrying the topological charges $m_1 = m_2 = 1$ or $m_1 = m_2 = 2$ in their components, are constructed by means of numerical and approximate analytical methods. The analysis reveals *stability regions* for the vortex droplets (in broad and relatively narrow parameter regions for $m_{1,2} = 1$ and $m_{1,2} = 2$, respectively). The results provide the first example of stable 3D self-trapped states with the double vorticity ($m_{1,2} = 2$), in any physical setting. The stable modes are shaped as flat-top ones, with the space between the inner hole, induced by the vorticity, and the outer boundary filled by a nearly constant density. On the other hand, all modes with *hidden vorticity*, i.e., topological charges of the two components $m_1 = -m_2 = 1$, are unstable. The stability of the droplets with $m_{1,2} = 1$ against splitting (which is the main scenario of possible instability) is explained by estimating analytically the energy of the split and un-split states. The predicted results may be implemented, exploiting dilute quantum droplets in mixtures of Bose-Einstein condensates.

*PhySH Subject Headings: Solitons; Bogoliubov-de Gennes equations; Superfluids; Mixtures of atomic and/or molecular quantum gases*

## 1. Introduction

The possibility of creation of three-dimensional (3D) solitons in nonlinear media was first proposed about 40 year ago [1]. It remains a challenging problem that connects diverse areas of physics and keeps drawing much interest up to this day [2-9]. A commonly known obstacle preventing the creation of 3D solitons in experiments is the collapse instability [10] of 2D and 3D states based on the ubiquitous cubic nonlinearity, which represents the Kerr self-focusing in optics [11] or attractive contact interactions in atomic gases cooled down to a Bose-Einstein condensate (BEC) [12]. Localized modes created in previous experiments may seem as nearly isotropic 3D fundamental (zero-vorticity) solitons – in particular, matter-wave solitons in BEC [13] – but their self-trapping is provided by the nonlinearity only in one direction, while in the others it is given by an external confining potential. In optics, the free-space collapse may be arrested using nonlinearities of other types, such as quadratic, composite focusing-defocusing, or time-delayed ones. However, severe experimental restrictions have led, thus far, only to the realization of effectively 2D stable self-trapping in these settings [14-16].

A still more challenging objective is the creation of stable 3D solitons with embedded vorticity (topological charge), $m$. While it was predicted that composite focusing-defocusing nonlinearities may readily create stable 2D vortex solitons with $m \geq 1$ [17], so far experiments were solely able to exhibit details of the splitting instability of vortex-ring solitons in the effectively 2D setting [18]. Only transient stabilization of a vortex ring with $m = 1$ under the additional action of three-photon absorption was demonstrated in a bulk optical medium with saturable nonlinearity [19]. In BEC, stability of 2D vortex solitons in free space was recently predicted, at least up to $m = 5$, in a polariton-type system including two components of the mean-field (MF) wave function coupled by a microwave field [20]. Concerning 3D systems in free space, the only prediction is a stability region for solitons supported by the focusing-defocusing cubic-quintic nonlinearity in the form of a vortex torus with $m=1$ [21] (the stability of 3D solitons with $m=0$ in the same model is completely obvious [22]). The above discussion focuses on self-trapping in uniform media/free space. The use of optical waveguides in the form of quasi-two-dimensional arrays, burnt in a bulk sample, makes it essentially easier to create quasi-discrete "light bullets" with embedded vorticity [23].

A breakthrough in the theoretical and experimental study of stable self-trapping in the 3D geometry has taken place in the course of the last two years. Works [24,25] theoretically proposed the possibility to create stable *droplets* in binary superfluids with an extremely low density, assuming intrinsic self-repulsion in each component, and dominating attraction between them. The collapse of the system, which is driven by the attractive interaction in 3D [10], is arrested by the beyond-MF correction to the energy resulting from zero-point quantum fluctuations around the MF state. The latter was derived by Lee, Huang, and Yang (LHY) in 1957 [26], and extended to binary mixtures in Ref. [27]. At low energies, an effective description of the system is provided by the addition of a repulsive quartic term to each Gross-Pitaevskii equation (GPE) describing the two-component system. Modifications of the system incorporating a linear Rabi coupling between the two components [28], or a linear spin-orbit coupling between them [29] were recently proposed as well. The use of binary condensates is a crucial condition for the realization of this approach, as the combination of the attractive cross-interaction, which is necessary for the self-trapping of 3D solitons, with the intrinsic self-repulsion induced by the LHY effect, which secures the stabilization against the collapse, is not possible in a single-component superfluid with only contact inter-atomic interactions. On the other hand, a single-component condensate made of dipolar atoms is sufficient for the creation of multidimensional solitons if the attraction is provided by long-range dipole-dipole forces, while the contact interaction, including the LHY term, remains repulsive [30-32].

The above-mentioned prediction was quickly followed by its experimental implementation. First, droplets were created in single-component dipolar condensates of $^{164}$Dy [30] and $^{166}$Er [31]. Due to the nature of the dipole-dipole interactions, the resulting droplets feature strong anisotropy. In addition to the experimental work, various aspects of the dynamics of droplets in dipolar superfluids were analyzed theoretically [32]. A possibility of the creation of dipolar droplets with embedded vorticity was recently considered too [33]. Like in many other models, all such vortex states are *completely unstable*, i.e., the use of dipolar condensates does not offer a solution to the problem of the creation of stable vortex solitons.

The creation of 3D zero-vorticity droplets, supported by the contact interactions, including the LHY effect, in binary mixtures of different internal states of $^{39}$K condensates was demonstrated very recently [34-36]. In these experiments, the necessary relation between the repulsive and attractive intra- and inter-component interactions is provided by suitable Feshbach resonances [37]. The experiments were performed both in the presence of confining potentials acting in one or two directions, and in free space.

The advent of this completely new technique for the realization of stable self-trapping in 3D ultra-dilute superfluids suggests the possibility to create stable droplets with *embedded vorticity*, which may assume both unitary and multiple values, $m=1$ and $m\geq 2$. The development of appropriate methods to achieve this goal is the subject of the present work. The possibility of the creation of such states is quite significant because in single-component dipolar condensates with LHY corrections only unstable vortex solitons may exist [33]. Furthermore, the only previously published prediction for the creation of stable 3D vortex rings with $m=1$ in free space relied on the 3D nonlinear Schrödinger equation (NLSE) with cubic-quintic nonlinearities. The latter is not accessible to BEC experiments, and in optics it has only been implemented in an effectively 2D setting [15]. We explicitly predict broad stability regions for 3D vortex rings with $m=1$ and $2$. This constitutes the first prediction of stable vortex solitons with a *multiple topological charge* in 3D. Note that stable modes with $m>2$ can be found too, but in a region which may be hard to access experimentally. While accurate results are produced by means of systematic numerical calculations, we also develop an analytical approximation to explain the stability of the 3D vortex rings against splitting, based on energy estimates for the un-split and split states. The analytical approach may be applied as well to a broad class of models in which 3D (or 2D) vortex solitons with "flat-top" (nearly constant-internal-density) profiles are created by competing nonlinearities. The theoretical results reported here suggest the possibility to create stable 3D droplets of swirling superfluids in current experiments [34-36]. To this end, vorticity should be imprinted onto binary droplets with $m=0$ via rotation of the system or by exploiting a twisted laser beam [38].

## 2. The model and analytical results

We model the binary superfluids using the system of coupled GPEs for the two-component wave function $\psi_{1,2}$, which have been extended to include the LHY quartic repulsion terms, in addition to the usual mean-field cubic ones. Equal scattering lengths $a$ of the contact interactions in both component are assumed here. We do not expect deviations from this condition to cause essential changes in the results. In the scaled form, the equations read [24,25]

$$i\frac{\partial \psi_1}{\partial t} = -\frac{1}{2}\nabla^2 \psi_1 + (|\psi_1|^2 + g_{\text{LHY}}|\psi_1|^3)\psi_1 - g|\psi_2|^2 \psi_1,$$
$$i\frac{\partial \psi_2}{\partial t} = -\frac{1}{2}\nabla^2 \psi_2 + (|\psi_2|^2 + g_{\text{LHY}}|\psi_2|^3)\psi_2 - g|\psi_1|^2 \psi_2, \tag{1}$$

where $\nabla^2 = \partial^2/\partial x^2 + \partial^2/\partial y^2 + \partial^2/\partial z^2$, the strength of the cubic self-repulsion is scaled to be 1, while $g>0$ and $g_{\text{LHY}} \approx (128/3)\sqrt{2/\pi}a^{3/2}$ are the relative strengths of the cross-attraction and additional LHY repulsion, respectively, $a$ being the intra-component scattering length. The extended GPE system (1) is valid for the description of relatively broad spatial patterns. This condition holds for the vortex droplets produced below because the diameter of their inner hole, induced by the vorticity, remains larger than the characteristic length associated to the most energetic Bogoliubov branch of the spectrum. This is the condition for which the validity of the equations was established [24,25]. It is relevant to mention that a more general treatment of the system including the LHY effect in the dilute binary condensate is possible, without resorting to the perturbative form of the GPEs, but rather running Monte-Carlo simulations of a multi-particle quantum system [39].

In fact, $g$ is the single irreducible parameter of the system, as additional rescaling $\{g_{\text{LHY}}\psi_{1,2}, g_{\text{LHY}}^{-2}t, g_{\text{LHY}}^{-1}\mathbf{r}\} \rightarrow \{(1/2)\psi'_{1,2}, t', \mathbf{r}'\}$, where $\mathbf{r}=\{x,y,z\}$, allows one to set $g_{\text{LHY}}\equiv 1/2$, all numerical results being presented below for this value. In general, we use the cross-attraction strength $g=1.75$. Although variations of the latter parameter cannot be absorbed by rescaling, they do not affect the results significantly. Equation (1) conserves two norms $N_{1,2}$, the total energy $E$, and the total linear $\mathbf{P}$ and angular $\mathbf{M}$ momenta, the $z$-component of the latter written below too:

$$N = \iiint (|\psi_1|^2 + |\psi_2|^2) dx dy dz \equiv N_1 + N_2,$$
$$E = (1/2)\iiint [|\nabla\psi_1|^2 + |\nabla\psi_2|^2 + |\psi_1|^4 + |\psi_2|^4 + (4/5)g_{\text{LHY}}(|\psi_1|^5 + |\psi_2|^5) - 2g|\psi_1|^2|\psi_2|^2] dx dy dz, \tag{2}$$
$$M_z = i\sum_{n=1,2}\iiint [y(\partial\psi_n/\partial x) - x(\partial\psi_n/\partial y)] dx dy dz.$$

We search for stationary solutions for vortex droplets, with chemical potentials $\mu_{1,2}$ and integer topological charges $m_{1,2}$ of their components. In cylindrical coordinates $(\rho,\theta,z)$ they are written as $\psi_{1,2} = u_{1,2}(\rho,z)\exp(im_{1,2}\theta - i\mu_{1,2}t)$, where the real wave functions $u_{1,2}$ obey the equations

$$\mu u_{1,2} + \frac{1}{2}\left(\frac{\partial^2}{\partial \rho^2} + \frac{1}{\rho}\frac{\partial}{\partial \rho} - \frac{m_{1,2}^2}{\rho^2} + \frac{\partial^2}{\partial z^2}\right)u_{1,2} - (u_{1,2}^2 + g_{\text{LHY}}u_{1,2}^3)u_{1,2} + gu_{2,1}^2 u_{1,2} = 0. \tag{3}$$

As follows from Eq. (3), the angular momentum of the stationary vortex mode is $M_z = N_1 m_1 + N_2 m_2$. In this work, we focus on the basic configuration, with $\mu_1 = \mu_2$ and $m_1 = \pm m_2$ corresponding to identical stationary components $u_1 = u_2$ of the binary condensate.

In this case, Eq. (3) gives rise to localized modes under the condition of $g>1$. In the limit of $g-1 \to +0$, the amplitude and chemical potential of the modes decrease as $A \sim g-1$, $\mu \sim (g-1)^3$, while their radial and axial sizes and norm grow as $R, Z \sim (g-1)^{-3/2}$, $N \sim (g-1)^{-5/2}$. The same asymptotic relations, with $(g-1)$ replaced by $g$, remain valid in the opposite limit, $g \to \infty$. On the other hand, for fixed $g$, the asymptotic behavior corresponding to both $g_{\mathrm{LHY}} \to +0$ and $g_{\mathrm{LHY}} \to \infty$ is $\mu \sim g_{\mathrm{LHY}}^{-2}$, $A \sim g_{\mathrm{LHY}}^{-1}$, $R, Z \sim g_{\mathrm{LHY}}$, $N \sim g_{\mathrm{LHY}}$.

In the case of equal vorticities of the components, $m_1 = m_2 \equiv m$, the total angular momentum is $Nm$, while for the vortex droplets with $m_1 = -m_2$ it is zero, hence modes of this type are called *hidden-vorticity* states [40]. While stationary shapes of explicit- and hidden-vorticity states are identical, their stability is drastically different. As shown below, the latter are completely unstable in the present system.

In Figs. 1 and 2, we show numerically that stable vortex modes always feature a *flat-top* profile, with a nearly constant amplitude of the wave function in a broad area of the $(\rho, z)$ plane – in fact, between the inner hole induced by the vorticity and an outer boundary of the droplet. The limit value of the nearly constant wave function with amplitude $A_{\mathrm{co}}$, filling the droplet, corresponds to the *cutoff* chemical potential, $\mu_{\mathrm{co}}$, below which the 3D modes do not exist. Both $A_{\mathrm{co}}$ and $\mu_{\mathrm{co}}$ can be found exactly. To this end, we notice that, in the limit of $\mu - \mu_{\mathrm{co}} \to +0$, the self-trapped modes become extremely broad, hence sufficiently far from the center the radial equation (3) for $u_{1,2}(\rho) \equiv u(\rho)$ becomes quasi-one-dimensional:

$$\mu u + (1/2) d^2 u / d\rho^2 + (g-1) u^3 - g_{\mathrm{LHY}} u^4 = 0. \tag{4a}$$

This equation can be derived from a formal Hamiltonian,

$$\mathcal{H} = (1/2)(du/d\rho)^2 + (1/2)\mu u^2 + (1/4)(g-1)u^4 - (1/5) g_{\mathrm{LHY}} u^5, \tag{4b}$$

where localized solutions, with $u(\rho = \infty) = 0$, correspond to $\mathcal{H} = 0$. Further, the asymptotically constant (flat) state implies setting $d^2 u / d\rho^2 = du/d\rho = 0$ and $\mathcal{H} = 0$ in Eqs. (4), which leads to a system of two algebraic equations for $\mu_{\mathrm{co}}$ and $A_{\mathrm{co}}$. Its solution is

$$\mu_{\mathrm{co}} = -(25/216)(g-1)^3 g_{\mathrm{LHY}}^{-2}, \quad A_{\mathrm{co}} = (5/6)(g-1) g_{\mathrm{LHY}}^{-1}. \tag{5}$$

This method allows one to find the asymptotic values of the chemical potential and field amplitude in any model producing flat-top states, including the 2D case.

The radius of the hole at the center of the vortex droplet may be evaluated by means of the Thomas-Fermi approximation (TFA), which is applicable to vortices in media with repulsive nonlinearity [42]. In the framework of the TFA, one drops derivatives in Eq. (3), arriving at the simplified equation: either $u=0$ or $(g-1)u^2 - g_{\mathrm{LHY}} u^3 = m^2/2\rho^2 - \mu$. In the core of the vortex, $u(\rho)$ must vanish at $\rho \to 0$. As a solution of the latter TFA equation, $u(\rho)$ can follow this trend, decreasing from asymptotic value (5) up to $u_{\min} = (2/3)(g-1)g_{\mathrm{LHY}}^{-1}$, attained at $\rho^2 = \rho_{\min}^2 = (108/7) m^2 g_{\mathrm{LHY}}^2 (g-1)^{-3}$. With the further decrease of $\rho$ the TFA solution has no other option but jumping down to $u=0$. Thus, $\rho_{\min}$ is the TFA prediction for the radius of the central hole. In particular, for the parameters of Fig. 2(b), $\rho_{\min} \approx 3$ is qualitatively consistent with the figure, and the prediction of $\rho_{\min} \sim m$ is consistent with Figs. 4 and 5.

The quasi-one-dimensional approximation based on Eq. (4b) with $\mathcal{H} = 0$ can be used to calculate the surface-energy density for the outer and inner boundaries of the two-component droplet: $\sigma = \int_0^{A_{\mathrm{co}}} (du/d\rho) du \approx 0.045 (g-1)^{7/2} g_{\mathrm{LHY}}^{-3}$, where identity $d\rho \equiv (du/d\rho)^{-1} du$ is used and the numerical factor is produced by the integral. One can derive a condition securing the *energy stability* of the droplet against the fission, taking into account the surface energy, the droplet's intrinsic vortex energy with density $\rho^{-2} |\partial \psi / \partial \theta|^2$, and the kinetic energy of two splinters produced by possible fission of the droplet with $m=1$. Momenta of the splinters, which give rise to the kinetic energy, are determined by the conservation of the original intrinsic angular momentum (spin) of the vortex. The latter is converted into the orbital angular momentum of the separating splinters. Thus, assuming a spherical shape of the droplet, with outer radius $R$ determined by the half-norm and nearly-constant density $A_{\text{co}}^2$, $N/2 = (4\pi/3)R^3 A_{\text{co}}^2$, and the central hole of radius $\rho_{\min}$, the condition guaranteeing that the total energy of the original vortex droplet is *smalle*r than the energy of the splinter pair takes the following form for large $N$:

$$(N/2)^{1/3} > \left(\frac{4\pi}{3}\right)^{1/3} \frac{A_{\text{co}}^{2/3}}{2^{1/3}-1} \left\{ \frac{A_{\text{co}}^2}{3\sigma} \left[ \ln\left(\frac{3N}{8\pi A_{\text{co}}^2 \rho_{\min}^3}\right) - 2^{2/3} \right] + \rho_{\min} \right\}. \tag{6}$$

Note that this condition takes into account the fact that the spherical splinters have only an outer surface, but no inner one, unlike the original doughnut-shaped vortex. In Eq. (6), the left-hand side represents the increase of the overall surface energy due to the fission, the logarithmic term on the right-hand side represents the original vortex energy (it is the dominant term on the right side for sufficiently large $N$), the next negative term is the opposite contribution to the energy balance from the splinters' kinetic energy, and the last term accounts for the surface energy of the inner hole in the original vortex droplet. While the approximation given by Eq. (6) is not accurate enough for detailed comparison with numerical data, it readily explains the stabilization of the vortex droplet for $N$ large enough, for this model and others with the flat-top shape of vortex solitons. On the other hand, for the hidden-vorticity mode (zero total angular momentum), the kinetic energy of the splinters does not appear in the energy balance, as there is no source for it, which makes the hidden-vorticity state more prone to the fission instability.

## 3. Numerical results

Families of numerically found zero-vorticity and vortex droplets for $m_1 = m_2 = 0, 1, 2$, including their stability, are characterized by dependences of the norm vs. $\mu_1 = \mu_2$ shown in Fig. 1(a). In particular, the stability of the fundamental modes with $m_1 = m_2 = 0$ exactly follows the Vakhitov-Kolokolov (VK) criterion, $dN/d\mu < 0$, which is a necessary condition for the stability of self-trapped modes supported by any attractive nonlinearity, irrespective of the spatial dimension [41,10]. Note that the branches of $N(\mu)$ curves with the norm diverging at $\mu_{1,2} \to 0$, which corresponds to the decaying amplitude $A_{1,2} \equiv \max\{u_{1,2}(x)\}$ and diverging width [see Fig. 1(c)], are definitely unstable according to the VK criterion. The presence of these branches is a characteristic feature of the 3D setting. On the other hand, the divergence of the norm at $\mu_{1,2} \to \mu_{\text{co}}$, where the cutoff value $\mu_{\text{co}}$ [dashed line in Fig. 1(a)] is exactly predicted by Eq. (5), implies the expansion of the above-mentioned flat-top states, filled by the wave function with the asymptotically constant value, which is also exactly predicted by Eq. (5). In particular, it does not depend on $m_{1,2}$. A typical example is displayed in Fig. 2(b). In agreement with the above analysis, the outer radius of the droplets grows with $N$ as $(3N/4\pi A_{\text{co}}^2)^{1/3}$, while the radius of the inner hole remains constant. An essential property of the 3D modes is that they exist with the norm exceeding a minimum value, $N_{\min}$, similar to what is known about other 3D models with competing nonlinearities [21]. The dependence of the droplet's energy on the norm shows a typical cusp-like shape for all values of $m_{1,2}$ [Fig. 1(b)]. Naturally, for fixed $N$ the energy of the self-trapped modes grows with the increase of the vorticity, driving the modes towards instability. Nevertheless, nontrivial findings produced by the present analysis are *stability regions* for the vortex droplets with $m_{1,2} = 1$, and even for $m_{1,2} = 2$, see details below.

The rigorous stability analysis is based on the consideration of weakly perturbed solutions, $\psi_{1,2} = (u_{1,2} + \alpha_{1,2} e^{\lambda t + ik\theta} + \beta_{1,2}^* e^{\lambda^* t - ik\theta}) e^{im_{1,2}\theta - i\mu_{1,2} t}$, where $\alpha_{1,2}, \beta_{1,2}$ are eigenmodes of small perturbations, $\lambda \equiv \lambda_r + i\lambda_i$ is the corresponding instability growth rate, which may be complex, and $k$ is an integer azimuthal perturbation index. The linearization of Eq. (1) for small perturbations leads to the corresponding Bogoliubov – de Gennes equations:

$$i\lambda\alpha_{1,2} = -\frac{1}{2}\left(\frac{\partial^2}{\partial\rho^2} + \frac{1}{\rho}\frac{\partial}{\partial\rho} - \frac{(m_{1,2}+k)^2}{\rho^2} + \frac{\partial^2}{\partial z^2}\right)\alpha_{1,2} + \left(2u_{1,2}^2 + \frac{5}{2}g_{\text{LHY}}u_{1,2}^3 - gu_{2,1}^2\right)\alpha_{1,2} +$$
$$\left(u_{1,2}^2 + \frac{3}{2}g_{\text{LHY}}u_{1,2}^3\right)\beta_{1,2} - gu_1u_2(\alpha_{2,1} + \beta_{2,1}) - \mu_{1,2}\alpha_{1,2},$$
$$i\lambda\beta_{1,2} = +\frac{1}{2}\left(\frac{\partial^2}{\partial\rho^2} + \frac{1}{\rho}\frac{\partial}{\partial\rho} - \frac{(m_{1,2}-k)^2}{\rho^2} + \frac{\partial^2}{\partial z^2}\right)\beta_{1,2} - \left(2u_{1,2}^2 + \frac{5}{2}g_{\text{LHY}}u_{1,2}^3 - gu_{2,1}^2\right)\beta_{1,2} -$$
$$\left(u_{1,2}^2 + \frac{3}{2}g_{\text{LHY}}u_{1,2}^3\right)\alpha_{1,2} + gu_1u_2(\alpha_{2,1} + \beta_{2,1}) + \mu_{1,2}\beta_{1,2},$$
(7)

which were solved numerically. Representative dependencies of real parts $\lambda_r$, i.e., the growth rate of the instability, on the chemical potential of the underlying stationary states with vorticities $m_{1,2}=1$ and $2$ are shown in Figs. 3(a) and 3(b). Typically, the imaginary part of the growth rate is comparable to its real part. Concerning the zero-vorticity states, the analysis demonstrates that (as mentioned above) the VK criterion is sufficient for their stability. However, this is not true for the vortex modes.

The central result of this work is that vortex droplets with $m_{1,2}=1$ and $2$ are *stable* in a certain interval, $\mu_{\text{co}}<\mu<\mu_{\text{st}}$, as shown in Fig. 3(c) for $m_{1,2}=1$. It is relevant to present the stability region vs. $g$, as the relative strength of the inter-component attraction may be adjusted experimentally by means of a Feshbach resonance [37]. Moreover, as mentioned above, $g$ is the only irreducible parameter in Eq. (1). Interestingly, the stability interval of $\mu$ expands with the increase of $g$. The stability domain in the $(g,N)$ plane is presented in Fig. 3(d). Note that, in terms of the norm, the stability domain for the vortex droplets is bounded only from below, i.e., droplets with $N>N_{\text{st}}$ are stable for arbitrarily large values of $N$. As seen in Fig. 3(d), the vortex droplets also exist in the adjacent interval of $N_{\text{min}}<N<N_{\text{st}}$, where they are unstable Note that the threshold value of the norm for the stability, $N_{\text{st}}$, rapidly decreases with increase of $g$. On the other hand, the divergence of $N_{\text{st}}$ at $g\to g_{\text{min}}\simeq 1.3$ implies that smaller values of $g$ cannot provide for stable equilibrium between the inter-component attraction and self-repulsion, including the LHY effect.

Figure 3(a) demonstrates that the most dangerous perturbation eigenmode for the droplets with $m_{1,2}=1$ has azimuthal index $k=2$. It tends to split the vortex mode in two fragments, see Fig. 4. The growth rate for this perturbation decreases and eventually vanishes with the increase of the norm of the vortex droplet, as it develops the flat-top (constant-inner-density) shape. The vortex becomes completely stable in a sufficiently broad range of chemical potentials, $\mu_{\text{co}}<\mu_{1,2}<\mu_{\text{st}}$.

The predicted stability of such states at moderate values of the norm is promising for their experimental realization. Although the droplets realized in Refs. [34-36] are not yet in a regime where the density shows a flat-top profile and vortex droplets are stable, other BEC mixtures could be exploited to observe them easier. For instance, in a $^{87}$Rb-$^{41}$K mixture [43], stable vortex droplets with densities ~5·10$^{15}$ atoms/cm³ and inner vortex diameter $\sim 0.5\,\mu$m might be created for parameters displayed in Fig. 3.

Representative scenarios of the evolution of unstable droplets with $m_{1,2}=1$, along with an example of the stable evolution, are displayed in Fig. 4 in the form of isosurfaces corresponding to a fixed density. In these cases, the evolution of unstable states was initiated by perturbations proportional to eigenmodes obtained from the numerical solution of Eq. (7). Unstable droplets split into several fragments, flying away tangentially to the initial ring in order to conserve the total angular momentum. The number of fragments is determined by the perturbation azimuthal index, $|k|$. For $k=1$, the original unstable vortex mode transforms into a stable zero-vorticity droplet, shedding off the initial angular momentum with emitted matter waves. If the instability is seeded by random initial perturbations, rather than by specially chosen eigenmodes, fission into two fragments is typically observed, because the corresponding instability mode has the largest $\lambda_r$ in Fig. 3(a). On the other hand, the 3D vortex rings which are predicted to be stable keep their shapes in direct simulations over indefinitely long time intervals (which actually exceed the experimental lifetime of the BECs), see the bottom row in Fig. 4.

The linear-stability analysis for the vortex droplets with $m_{1,2}=2$ shows that the spectrum of unstable perturbations includes larger values of azimuthal index $k$. The most destructive perturbations have $2 \leq k \leq 4$, depending on the chemical potential of the stationary state. A remarkable fact is that these 3D vortex rings, carrying a multiple topological charge, also have a stability domain clearly visible in Fig. 3(b), even if it is relatively narrow. This finding offers the first example of *stable* 3D self-trapped states with higher values of the topological charge in any matter-wave or optics setting. The stability is achieved for sufficiently broad states with a large norm. Figure 5 shows that, due to the richer spectrum of unstable perturbation eigenmodes, unstable droplets with $m_{1,2}=2$ may split into a larger number of fragments than their counterparts with $m_{1,2}=1$. If the instability is seeded by random perturbation and $|\mu_{1,2}|$ is not too large, the double-vorticity droplet usually splits in four fragments, in accordance with the prediction provided by Fig. 3(b). A typical example of the evolution of a stable vortex droplet with $m_{1,2}=2$, perturbed by broadband noise, is displayed in the bottom row of Fig. 5.

The existence of stable droplets with even higher vorticities is expected too, but, due to the rapid shrinkage of the stability domain (in terms of $\mu$) with the increase of the vorticity, such states may stabilize at values of the norm too large to be realized experimentally. Finally, the hidden-vorticity modes with $m_1 = -m_2 = 1$, which have characteristics and shapes identical to those displayed in Figs. 1 and 2, turn out to be unstable against fission in the parameter regime that may be relevant to experiments. This result is obtained through both the calculation of the stability eigenvalues and direct simulations. Fission dynamics of such states is illustrated in Fig. 6. In contrast to the case of $m_1 = m_2 = 1$, fragments resulting from the decay of the droplets with $m_1 = -m_2 = 1$ fly apart in the radial direction, rather than tangentially to the initial ring. This is due to the fact that the original state has zero angular momentum. Shapes of the $\psi_1$ and $\psi_2$ components remain close, but not identical, in the course of the fission.

## 4. Conclusions

In this work we have predicted a new scenario for the creation of self-trapped vortex rings, a type of 3D topological self-bound state never observed in any physical setting (even in effectively 2D geometries). We have predicted that these vortex rings should be stable in droplets of attractive two-component swirling superfluids. This system is described by a system of two coupled Gross-Pitaevskii equations with inter-component attraction and quartic beyond-mean-field repulsion due to the LHY (Lee-Huang-Yang) correction to the energies. Exploiting this model, we have constructed stationary states in the form of vortex rings with embedded topological charges $m_1 = m_2 = 1$ and $m_1 = m_2 = 2$ of the two components. These solutions represent 3D superfluid droplets with intrinsic vorticity. The latter can be imprinted onto the droplets either via rotation of the system or exploiting a twisted laser beam. Our main result is the discovery of *stability regions* for the vortex droplets, which are broad for $m_{1,2}=1$ and narrower for $m_{1,2}=2$. They constitute the first stable 3D states with double intrinsic vorticity predicted in any physical context. The stable modes feature flat-top shapes, i.e., ones with a nearly constant density filling the space between the inner hole, induced by the vorticity, and the outer boundary of the droplet. In addition, we have found that the droplets with *hidden vorticity*, i.e., topological charges $m_1 = -m_2 = 1$ in the two components, are completely unstable in the relevant parameter region. We investigate the stability of the vortex rings with $m_{1,2}=1$ against fission, using both systematic numerical studies and analytical estimates for the energy of the system. The analytical method can be applied to a broad class of models generating flat-top states, in 3D and 2D settings alike. Other analytical methods developed in this work, such as the prediction of the density filling the flat-top area and the radius of the vorticity-induced hole, may also be applied to a broad class of models.

The results reported here may stimulate the creation of 3D vortex droplets using mixtures of Bose-Einstein condensates, where droplets were recently demonstrated.

The present analysis may be developed in other directions. In particular, a natural objective is to consider interactions (collisions) of zero-vorticity and vortical droplets (in particular, of droplets with opposite topological charges, $m_{1,2}=1$ and $m_{1,2}=-1$). A challenging objective is to explore whether stable *hopfions*, i.e., vortex tori with an intrinsic twist, may exist. They represents the second independent topological charge of the self-trapped mode [44].

Acknowledgments. We gratefully acknowledge help of C. Milian in the development of the codes for the stability analysis, and valuable discussions with C. R. Cabrera, J. Sanz and G. E. Astrakharchik. LT, LT and YVK acknowledge support from the Severo Ochoa Program (SEV-2015-0522) of the Government of Spain, Fundacio Cellex, Fundació Mir-Puig, Generalitat de Catalunya and CERCA. YVK acknowledges partial support by the program 1.4 of Presidium of RAS "Topical problems of low temperature physics". The work of BAM on this project was supported in part by the joint program in physics between NSF and Binational (US-Israel) Science Foundation through project No. 2015616, and by the Israel Science Foundation through Grant No. 1286/17. This author appreciates hospitality of ICFO-Institut de Ciencies Fotoniques (Barcelona).

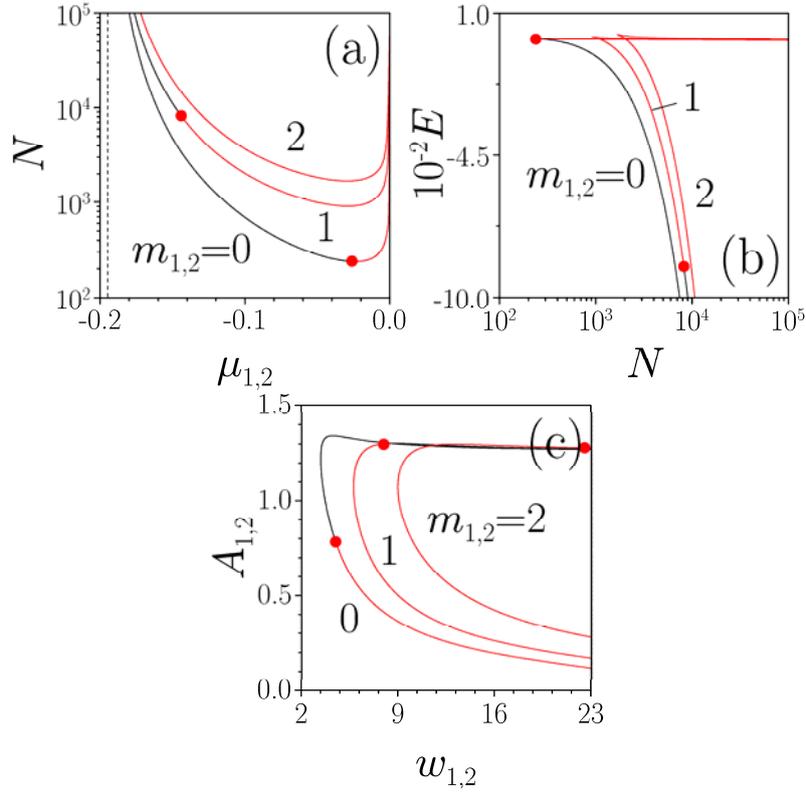

Figure 1. The norm versus the chemical potential (a), the energy versus the norm (b), and the amplitude of the wave function versus the width (c) for droplet modes with indicated values of the vorticity. Red and black branches correspond to unstable and stable states, respectively, with red dots separating stable and unstable segments. For $m_{1,2}=2$ a stability segment exists too, as shown in (c), but it is located at large values of $N$ outside of the regions displayed in (a) and (b). The dashed line in (a) designates the value of $\mu_{\text{co}}$, given by Eq. (5), below which localized modes do not exist. The results are displayed, here and in other figures, for $g_{\text{LHY}}=0.50$ and $g=1.75$.

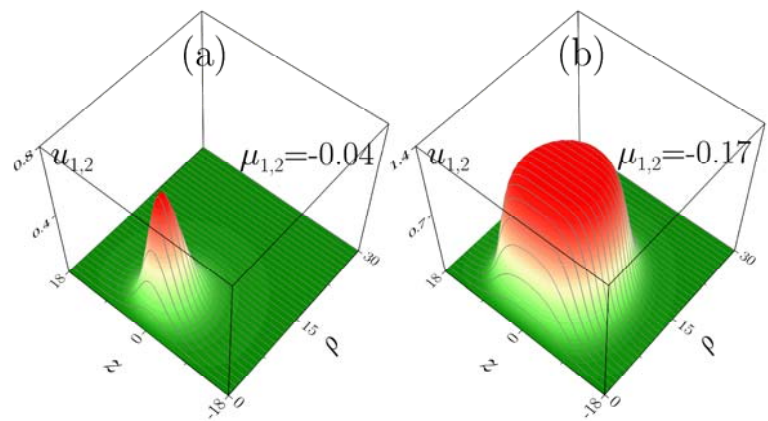

Figure 2. Typical profiles of unstable sharp (a) and stable flat-top (b) vortex droplets with $m_{1,2}=1$ for values of $\mu$ indicated in the panels. Here $g_{\mathrm{LHY}}=0.50$ and $g=1.75$.

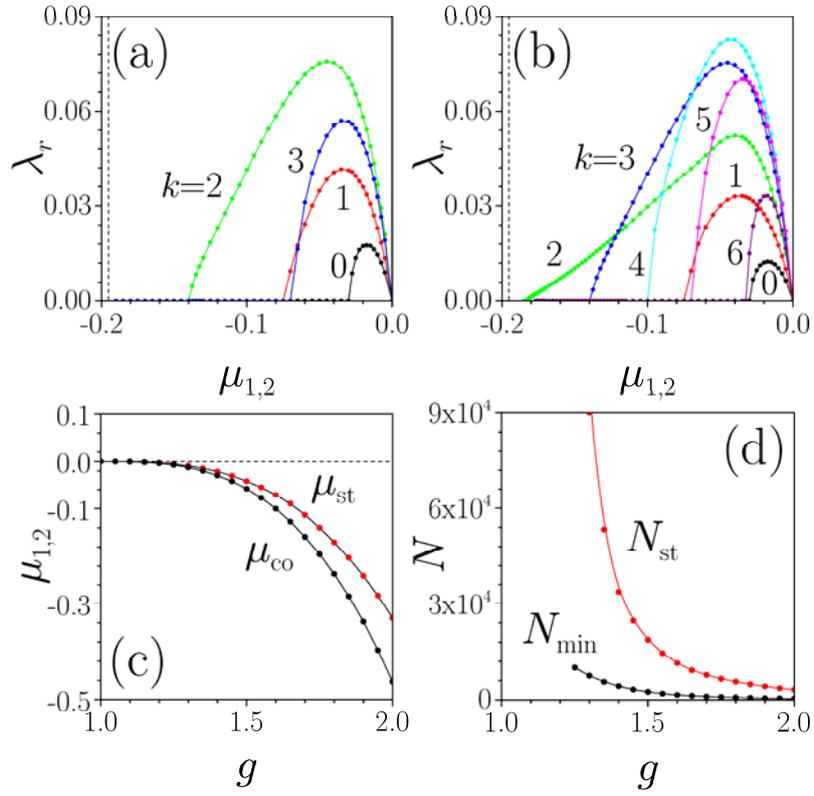

Figure 3. The real part of the instability growth rate versus the chemical potential of the stationary droplet, $\mu$, at values of the perturbation azimuthal index $k$ indicated in the panels, for $m_{1,2}=1$ (a) and $m_{1,2}=2$ (b) at $g_{\mathrm{LHY}}=0.50$ and $g=1.75$. The vertical dashed line designates the cutoff value of $\mu$ given by Eq. (5). Thus, *stability windows* are clearly observed in both panels (a) and (b), even if the window is rather narrow in (b). Panel (c) shows the cutoff value, $\mu_{\mathrm{co}}$, and the stability boundary, $\mu_{\mathrm{st}}$, versus the relative strength, $g$, of the inter-component attraction, for the vortex droplets with $m_{1,2}=1$. They are *stable* in the region of $\mu_{\mathrm{co}}<\mu<\mu_{\mathrm{st}}$. Panel (d) shows minimal norm $N_{\mathrm{min}}$ of vortex droplets and norm $N_{\mathrm{st}}$ above which such states become stable as functions of $g$. In (c) and (d) $g_{\mathrm{LHY}}=0.50$.

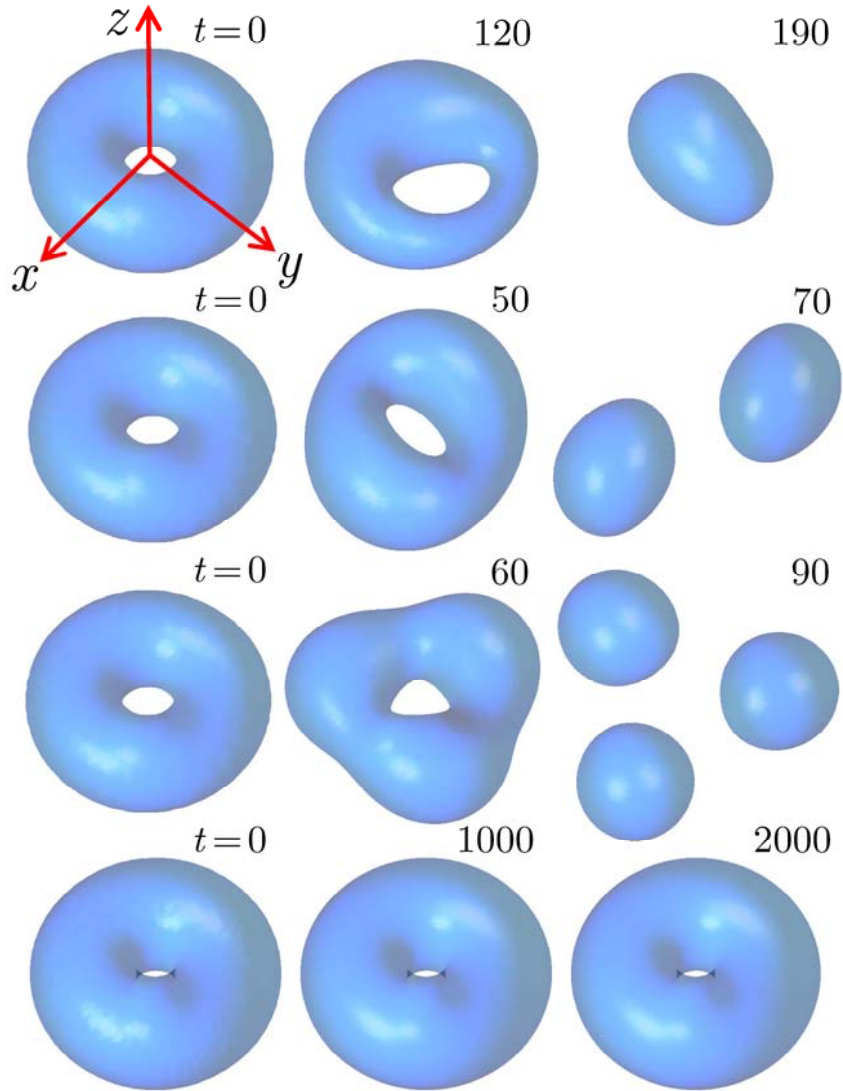

Figure 4. Isosurface density plots, at levels $|\psi_1|^2 = 0.1$ in the first three rows, and $|\psi_1|^2 = 0.5$ in the bottom one, showing unstable evolution of the perturbed vortex droplet with $m_{1,2} = 1$ and $\mu_{1,2} = -0.04$ (first to third rows), and stable evolution for $\mu_{1,2} = -0.16$ (the fourth row). Only the $\psi_1$ component is displayed. The breakup of the droplets in the first, second, and third rows is induced by small perturbations proportional to unstable eigenmodes produced by the solution of Eq. (7) with azimuthal indices $k = 1, 2, 3$, respectively. A broadband random perturbation is applied to the stable droplet in the bottom row. In all cases $g_{\mathrm{LHY}} = 0.50$ and $g = 1.75$.

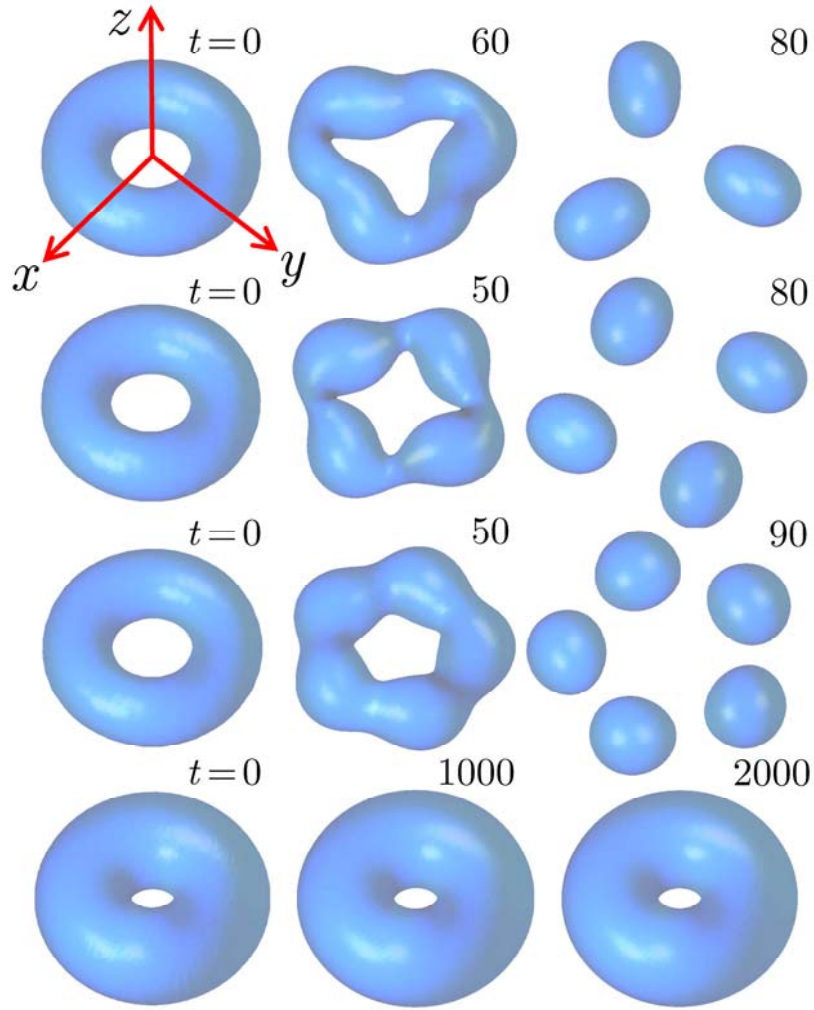

Figure 5. The same as in Fig. 4, but for droplets with intrinsic vorticity $m_{1,2}=2$ and chemical potentials $\mu_{1,2}=-0.04$ and $\mu_{1,2}=-0.183$ for the unstable state in the first three rows, and stable one in the bottom row, respectively. All the isosurfaces are plotted at $|\psi_1|^2=0.05$. The breakup of the droplet in the first, second, and third rows is initiated by perturbation eigenmodes corresponding to azimuthal indices $k=3,4,5$, respectively. In all cases $g_{\text{LHY}}=0.50$ and $g=1.75$

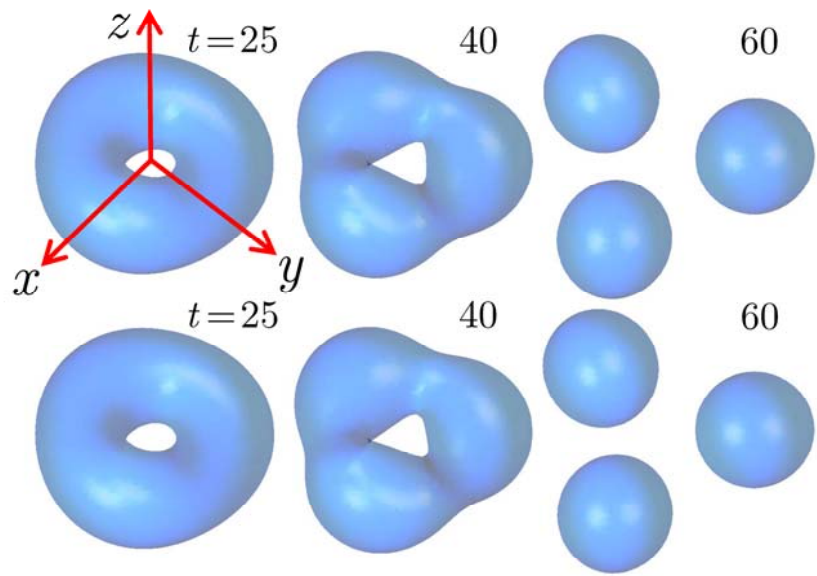

Figure 6. Isosurface density plots, at levels $|\psi_{1,2}|^2 = 0.1$ showing unstable evolution of the perturbed vortex droplet with $m_1 = +1$, $m_2 = -1$ and $\mu_{1,2} = -0.04$. Top row shows $\psi_1$ component, bottom row shows $\psi_2$ component. The breakup is induced by small perturbation with azimuthal index $k = 3$. In all cases $g_{\text{LHY}} = 0.50$ and $g = 1.75$.